\documentclass{cimento}

\usepackage{graphicx}  
\graphicspath{{./}}

\title{Charged-particle multiplicity with ALICE at the LHC}
\author{D.~Elia for the ALICE Collaboration}
\instlist{\inst{} INFN Bari, Italy}
\PACSes{\PACSit{12.38.Aw}{General properties of QCD (dynamics, confinement, etc)}
\PACSit{25.75.Dw}{Particle production (relativistic collisions)}}
\begin{document}

\maketitle

\begin{abstract}
The pseudorapidity density and multiplicity distributions of charged particles 
have been the first measurements carried out with the ALICE detector at the LHC.
After an introduction on the experiment and some details on the subdetectors relevant 
for these measurements, results from minimum bias proton-proton collisions
at 0.9, 2.36 and 7 TeV are presented. Comparisons with other measurements
and model predictions are also discussed. 
\end{abstract}

\section{ALICE experiment at the LHC}

ALICE is a general-purpose detector designed to measure the properties
of strongly interacting matter created in heavy-ion collisions at the CERN LHC.
Several features, such as low momentum cut-off and powerful tracking over a broad
momentum range, make it also an important contributor to the proton-proton LHC physics:
here ALICE aims both at setting the baseline for the understanding of the heavy-ion 
data and exploring the new energy domain \cite{ref:aliceppr2}.
\par
\begin{figure}
\centering
\includegraphics[scale=0.45]{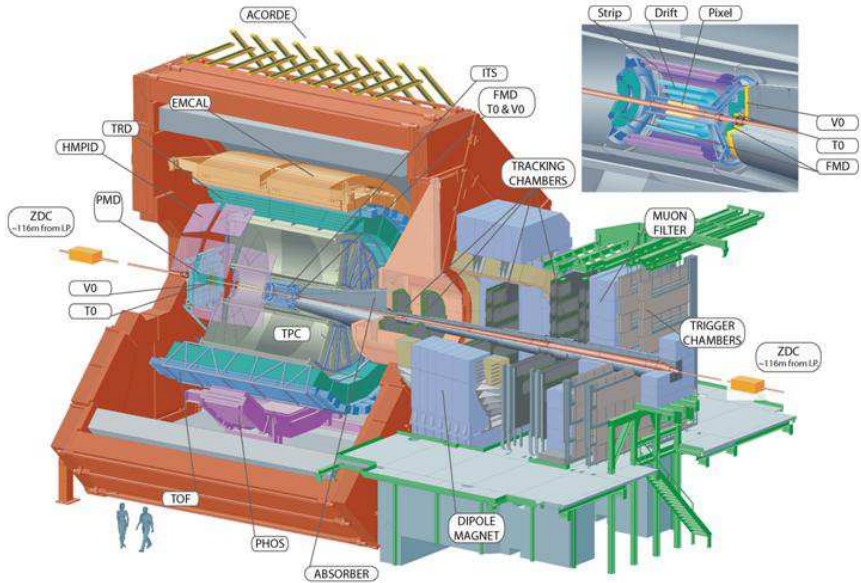}
\caption{General view of the ALICE detector and details of the innermost silicon layers.}
\label{ExperimentLayout}
\end{figure}

The ALICE experimental apparatus has been designed as a dedicated heavy-ion detector
optimized to measure a large variety of observables in very high multiplicity
environments with its performance able to resolve up to 8000 charged particles per unit of rapidity.
ALICE is able to detect and identify hadrons, leptons and photons over a wide range of momenta.
The whole detector, shown in Figure \ref{ExperimentLayout}, consists of a central
part (with acceptance $|\eta|$ $<$ 0.9) to detect hadrons, electrons and photons, 
a forward spectrometer to measure muons and additional smaller forward detectors for event 
cha\-racterization and triggering. In the following the Silicon Pixel Detector (SPD) 
is briefly described since it has been playing a key role for the first data
and the charged-particle multiplicity analysis: further details on the whole apparatus
can be found in \cite{ref:aliceppr1,ref:aliceapp}.

The SPD is the innermost part of the
ALICE vertex and tracking system. It consists of two cylindrical layers
placed at 3.9 and 7.6~cm average radii, equipped with hybrid silicon
pixels for a total of about 9.8 M individual cells of size 50~$\times$~425~$\mu$m$^2$.
It covers the pseudorapidity ranges $|\eta|$~$<$~2 and $|\eta|$~$<$~1.4
with the inner and outer layer respectively, for particles originating in the
centre of the detector. A fast signal (FastOR) is provided by each of the 1200 detector 
chips when at least one of its pixels is hit: 
all the FastOR signals are combined in a programmable logic unit 
which provides a level-0 trigger signal to the central trigger processor.
The total number of active channels during the data taking
was about 85\%. The detector alignment has been obtained using both cosmic-ray
tracks and tracks from proton-proton collisions \cite{ref:alicealign}.

\section{Tracklet reconstruction with the silicon pixels}

Besides its key role as a vertex detector, the SPD allows to determine the
number of charged primary particles in the central rapidity region.
The basic steps of the reconstruction algorithm are described in the following.
\par
First the position of the primary vertex is determined by correlating
hits in the two SPD layers, then using the reconstructed vertex position as 
the origin, tracklets candidates are reconstructed:
pairs of hits with one hit in each SPD layer are combined and the 
corresponding differences in azimuthal ($\Delta\varphi$, bending plane) and polar 
($\Delta\theta$, non-bending direction) angles are computed. 
Good tracklets are then selected as hit combinations satisfying a cut
on the sum of the squares of $\Delta\varphi$ and $\Delta\theta$ with a cut-off
of 80 mrad and 25 mrad respectively.
In case different candidates share a hit in one of the two layers, only the
combination with the smallest angular difference is taken as a good tracklet.
The cut imposed on $\Delta\varphi$ would reject charged particles with transverse momentum 
below 30 MeV/$c$; however, the effective transverse-momentum cut-off
due to absorption in the material is approximately 50 MeV/$c$.
\par
The number of primary charged particles is obtained from the number of tracklets 
corrected for the following effects: trigger and vertexing efficiency,
geometrical acceptance, detector and reconstruction efficiencies,
undetected particles below the transverse-momentum cut-off,
background from secondaries.
These effects and the corresponding corrections have been determined, using dedicated
Monte Carlo productions, as a function of the $z$-position of the primary vertex and 
the pseudorapidity of the tracklet. Further details on the reconstruction and
correction methods are given in \cite{ref:alicepp1,ref:alicepp2}.

\section{Pseudorapidity density and multiplicity distributions}

ALICE has measured the charged-particle pseudorapidity density distribution
with the very first low statistics sample of proton-proton collisions at a 
centre-of-mass energy $\sqrt{s}$=0.9 TeV, provided by the LHC at the end of 2009 \cite{ref:alicepp1}. 
Results presented in this contribution are based on a larger statistics
data sample at $\sqrt{s}$=0.9 TeV as well as the first higher energy
collisions at 2.36 and 7 TeV, all from the 2010 run.
Data at 0.9 and 7 TeV have been collected with a trigger requiring
a hit in either one of the trigger scintillator counters (VZERO) 
or in the SPD detector (i.e. at least one charged particle anywhere
in 8 units of pseudorapidity) in coincidence
with signals from the two beam-pickup counters (BPTX).
At 2.36 TeV the VZERO detector was turned off and the trigger required
at least one hit in the SPD in coincidence with the BPTX.
\par
\begin{figure}[htb]
\includegraphics[scale=0.33]{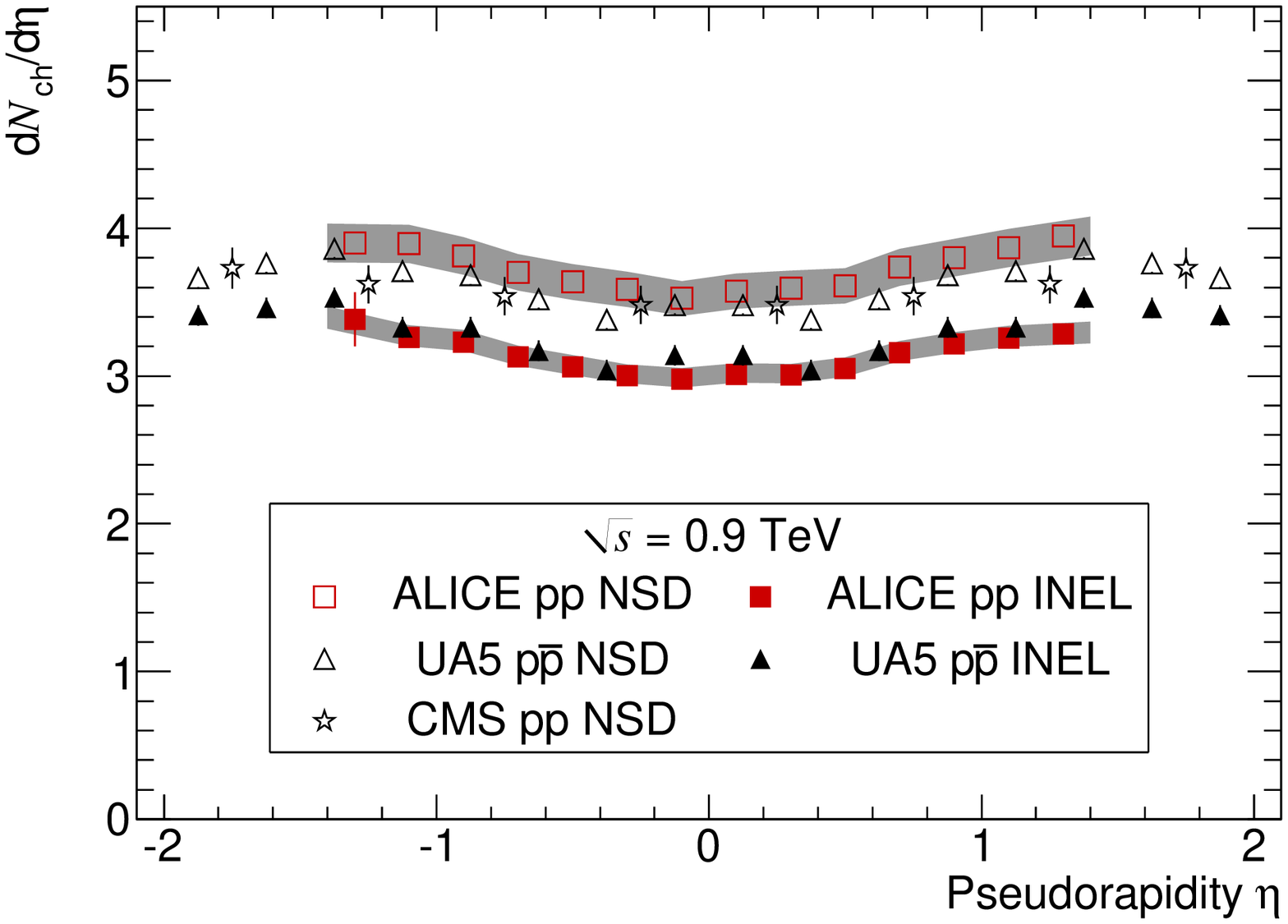}
\hspace{\fill}
\begin{minipage}[t]{110mm}
\includegraphics[scale=0.33]{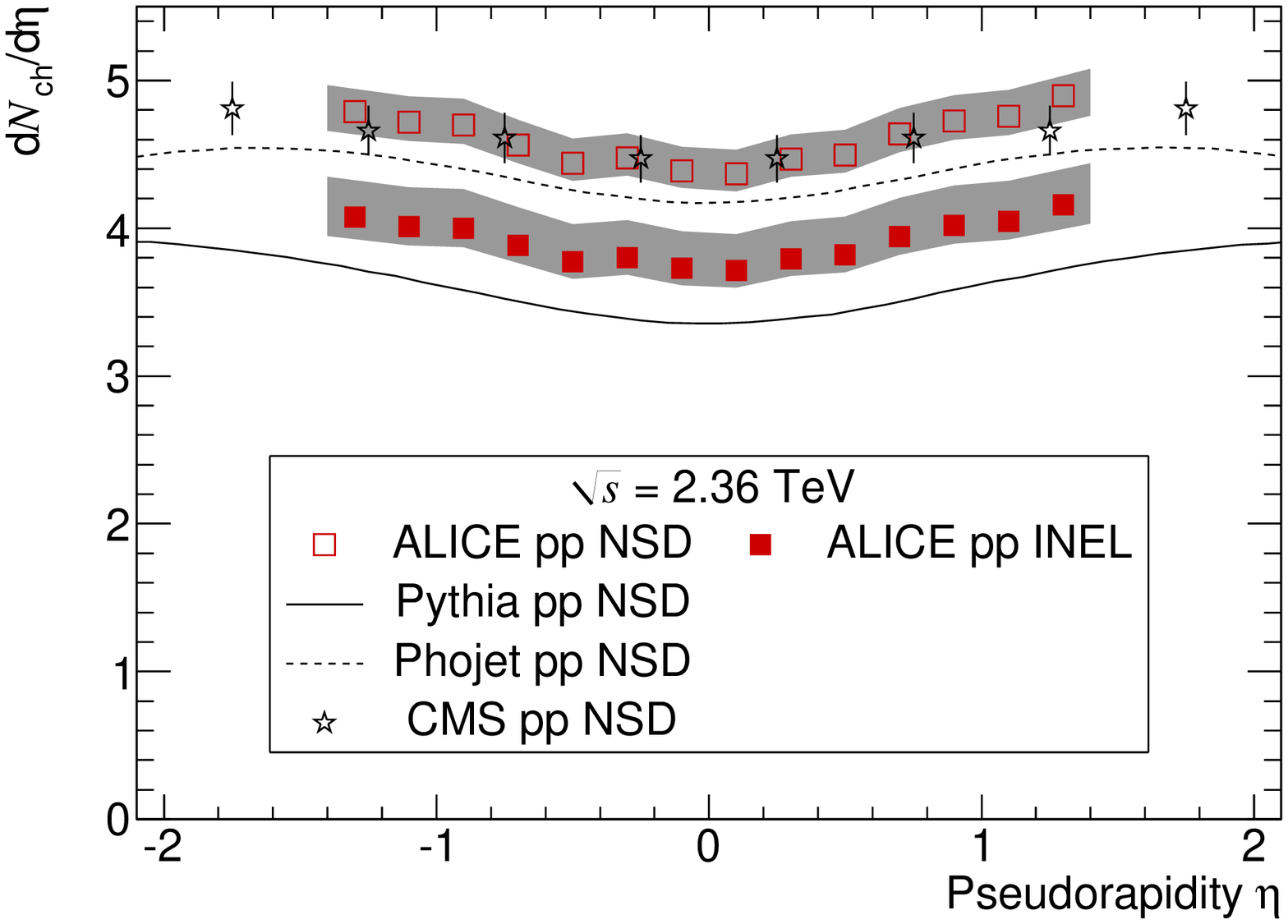}
\end{minipage}
\caption{Left panel: pseudorapidity density distributions for INEL and NSD 
interations at $\sqrt{s}$=0.9 TeV, compared to UA5 and CMS data. 
Righ panel: same for $\sqrt{s}$=2.36 TeV data, compared to CMS results and
predictions from PYTHIA and PHOJET \cite{ref:alicepp2}.}
\label{dndeta09236}
\end{figure}

Results at $\sqrt{s}$=0.9 and 2.36 TeV are based on 150000 and 40000 collision
events respectively and are given for two different normalizations:
inelastic (INEL), corresponding to the sum of all inelastic interactions,
and non-single-diffractive (NSD),
where the corrections for the two samples have been based on
previous experimental data and Monte Carlo simulations. 
In the left panel of Figure \ref{dndeta09236} the charged-particle pseudorapidity density
distribution for both INEL and NSD interactions at $\sqrt{s}$=0.9 TeV
are compared with p-$\overline{\rm p}$ data from the UA5 experiment \cite{ref:ua5} 
and with p-p NSD data from the CMS experiment \cite{ref:cms}.
The result is consistent both with UA5 and CMS data. 

The right panel of Figure \ref{dndeta09236} shows the measurement for
$\sqrt{s}$=2.36 TeV, compared to CMS NSD data \cite{ref:cms} and 
to PYTHIA tune D6T (109) \cite{ref:pythia} and PHOJET \cite{ref:phojet} calculations. 
The ALICE results for NSD collisions are consistent with CMS measurements,
syste\-matically above the PHOJET curve and significantly higher than the
PYTHIA tune D6T prediction. 
The pseudorapidity density measurements in the central region ($|\eta|$~$<$~0.5),
for both the event classes and at the two energies,
are summarized in Table \ref{tab:dndetavalues09236} together with UA5 and CMS
results and both model predictions.

\begin{table}[hbt]
  \caption{Charged-particle pseudorapidity densities measured by ALICE in p-p collisions
at $\sqrt{s}$=0.9 and 2.36 TeV for INEL and NSD event classes, compared with CMS and UA5
data as well as with PYTHIA tune D6T and PHOJET predictions. For ALICE the first error
is statistical and the second is systematic.}
  \label{tab:dndetavalues09236}
  \begin{tabular}{cccccc}
    \hline
      { }  & ALICE p-p        & CMS p-p                     &  UA5 p-$\overline{\rm p}$  &  PYTHIA  &  PHOJET   \\
    \hline
      {$\sqrt{s}$=0.9 TeV}  & { }   &  { }  &  {}  &  { }  &  { }  \\
    \hline
      INEL & 3.02 $\pm$ 0.01$^{+ 0.08}_{- 0.05}$  & { }                         &  3.09 $\pm$ 0.05   &  2.35      &  3.21     \\
      NSD  & 3.58 $\pm$ 0.01$^{+ 0.12}_{- 0.12}$  & 3.48 $\pm$ 0.02 $\pm$ 0.13  &  3.43 $\pm$ 0.05   &  2.85      &  3.67     \\
    \hline
      {$\sqrt{s}$=2.36 TeV}  & { }   &  { }  &  {}  &  { }  &  { }  \\
    \hline
      INEL & 3.77 $\pm$ 0.01$^{+ 0.25}_{- 0.12}$  & { }     &  { }                       &  2.81             &  3.76     \\
      NSD  & 4.43 $\pm$ 0.01$^{+ 0.17}_{- 0.12}$  & 4.47 $\pm$ 0.04 $\pm$ 0.16  &  { }   &  3.38             &  4.20     \\
    \hline
  \end{tabular}
\end{table}

The main contribution to the systematic uncertainties at 0.9 and 2.36 TeV comes
from the limited knowledge of cross section and kinematics of diffractive
processes. At 7 TeV, due to lack of experimental information about these processes,
a different event class has been defined by requiring at least one charged particle
in the pseudorapidity interval $|\eta|$~$<$~1 (INEL$>$0$_{|\eta|<1}$), minimizing
the model dependence of the corrections.
\par
\begin{figure}[hbt]
\centering
\includegraphics[scale=0.38]{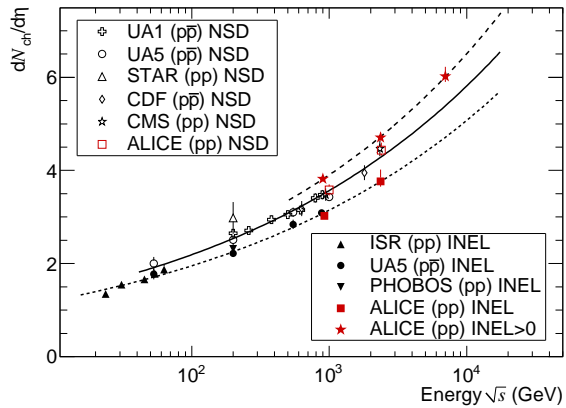}
\caption{Charged-particle pseudorapidity density in $|\eta|$~$<$~0.5 for
INEL and NSD event classes and in $|\eta|$~$<$~1 for the INEL$>$0$_{|\eta|<1}$
event class, as a function of the centre-of-mass energy. The lines indicate the fit
using a power-law dependence on energy \cite{ref:alicepp3}.}
\label{dndetavsenergy}
\end{figure}

Results at $\sqrt{s}$=7 TeV are based on 300000 p-p collisions. Data at lower energies 
have been re-analyzed in order to normalize the results to the same event class.
In Figure \ref{dndetavsenergy} the centre-of-mass energy dependence of the
charged-particle pseudorapidity density for the INEL$>$0$_{|\eta|<1}$ class
is compared to the evolution for the INEL and NSD classes at lower energies \cite{ref:alicepp3}. 

The observed trend is similar to that observed between 0.9 and 2.36 TeV for 
inelastic and non-single-diffractive events, then with a stronger increase
with the energy in the data with respect to the model predictions.  
In Table \ref{tab:dndetavalues092367} the pseudorapidity density values for $|\eta|$~$<$~1 
measured by ALICE at the three energies are reported and compared with predictions from 
different PYTHIA tunes (namely D6T (109), ATLAS-CSC (306) \cite{ref:pythiaatlas} and 
Perugia-0 (320) \cite{ref:pythiaperugia}) and from PHOJET.

\begin{table}[hbt]
  \caption{Charged-particle pseudorapidity densities in p-p collisions
at $\sqrt{s}$=0.9, 2.36 and 7 TeV measured by ALICE for INEL$>$0$_{|\eta|<1}$ event class, 
compared with different PYTHIA tunes and PHOJET predictions.}
  \label{tab:dndetavalues092367}
  \begin{tabular}{cccccc}
    \hline
      {Energy (TeV)}  & ALICE  & { }  &  PYTHIA  & { }  &  PHOJET   \\
      { }             &  { }   &  (109)  & (306) & (320) &   { }  \\
    \hline
      0.9  & 3.81 $\pm$ 0.01$^{+ 0.07}_{- 0.07}$  & 3.05 &  3.92  &  3.18  &  3.73     \\
      2.36 & 4.70 $\pm$ 0.01$^{+ 0.11}_{- 0.08}$  & 3.58 &  4.61  &  3.72  &  4.31     \\
      7    & 6.01 $\pm$ 0.01$^{+ 0.20}_{- 0.12}$  & 4.37 &  5.78  &  4.55  &  4.98     \\
    \hline
  \end{tabular}
\end{table}

\begin{figure}[htb]
\includegraphics[scale=0.33]{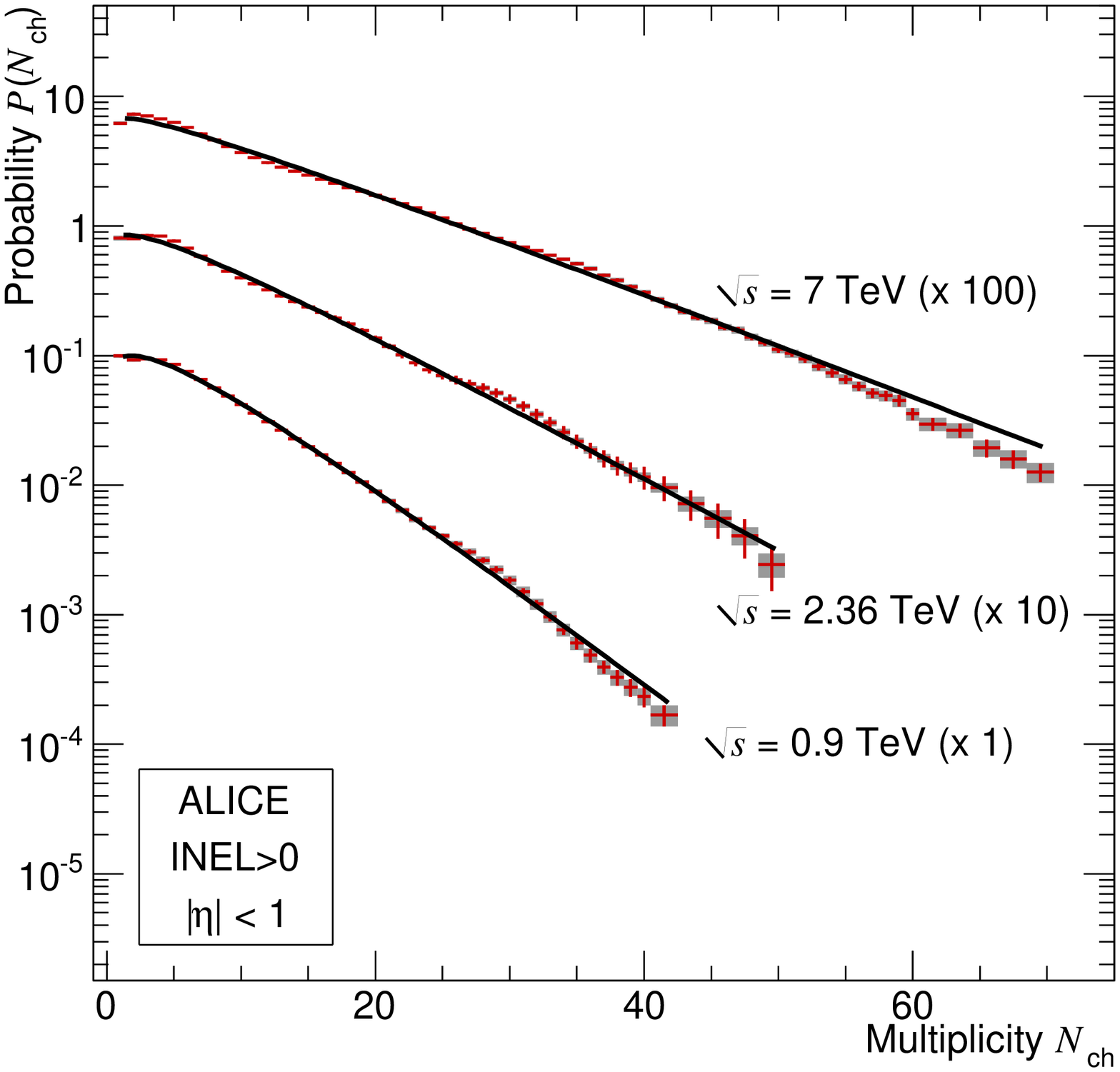}
\hspace{\fill}
\begin{minipage}[t]{110mm}
\includegraphics[scale=0.33]{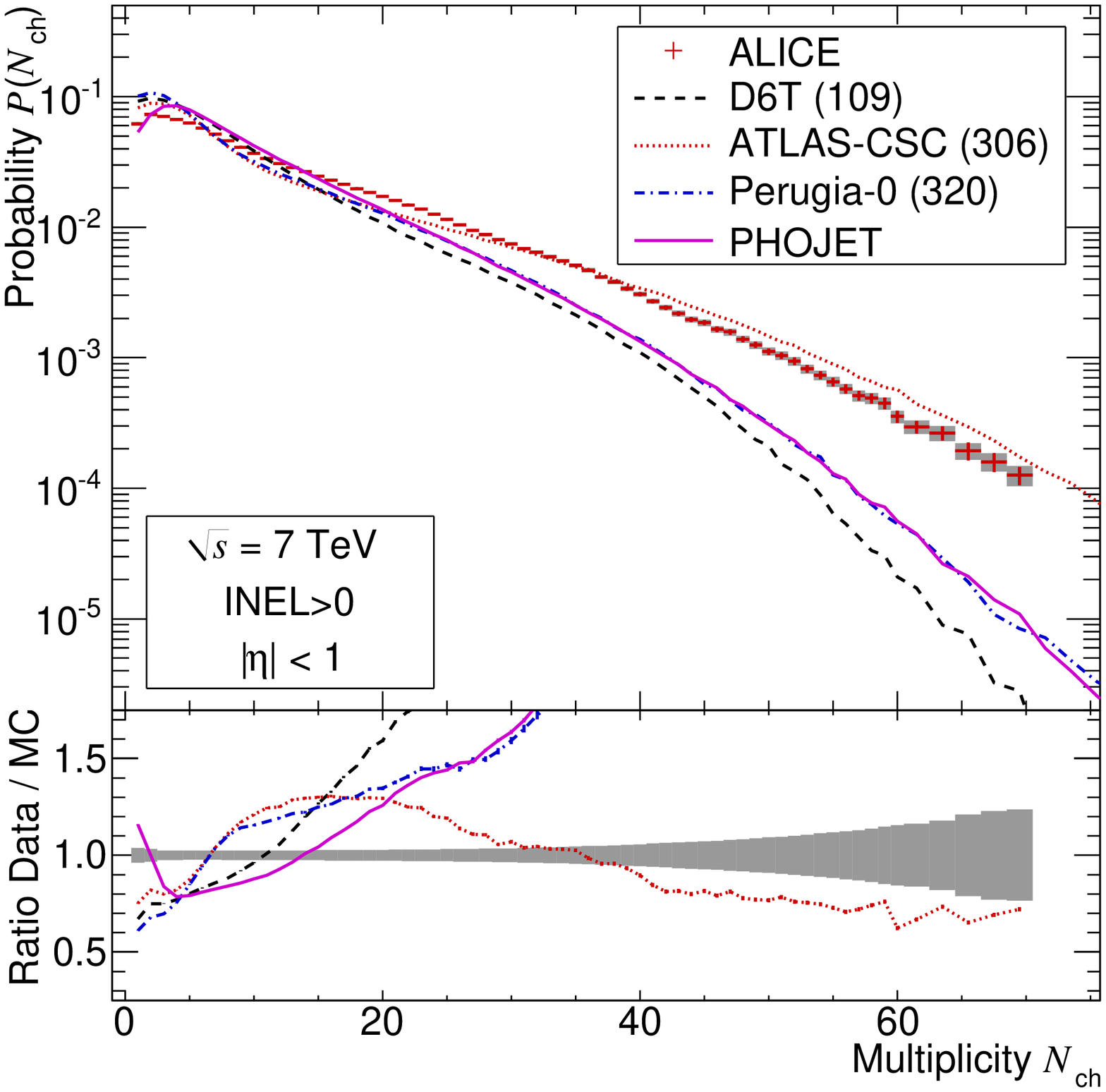}
\end{minipage}
\caption{Left panel: multiplicity distributions in $|\eta|$ $<$ 1 for 
the INEL$>$0$_{|\eta|<1}$ class at the three energies
with the NBD fits. The error bars represent the statistical errors
while the shaded areas represent the systematics.
Righ panel: data at $\sqrt{s}$=7 TeV compared with models \cite{ref:alicepp3}.}
\label{multiplicityspectra}
\end{figure}

An increase of 57.6\% is observed going from the 0.9 to the 7 TeV data, compared with an
increase of 47.6\% obtained from the closest model (PYTHIA ATLAS-CSC) \cite{ref:alicepp3}.
To better understand the increase from 0.9 to 2.36 TeV then to 7 TeV the multiplicity 
distributions for the INEL$>$0$_{|\eta|<1}$ event class have been studied as well.
In the left panel of Figure \ref{multiplicityspectra} the unfolded distributions
at 0.9 and 2.36 TeV are well described by the Negative Binomial Distribution (NBD),
while at 7 TeV the NBD fit slightly underestimates and overestimates the data at low
and at high multiplicity respectively.

The multiplicity distribution for the 7 TeV data is compared with
models in the right panel of Figure \ref{multiplicityspectra}: only PYTHIA
tune ATLAS-CSC is close to the data at high multiplicity ($N_{ch}$ $>$ 25),
even if it fails to reproduce the data in the intermediate region
(8 $<$ $N_{ch}$ $<$ 25). At low multiplicities a large spread between the
different models is observed, PHOJET being the lowest and PYTHIA tune Perugia-0
the highest \cite{ref:alicepp3}.

\section{Conclusions}

The charged-particle pseudorapidity density and multiplicity distributions
as measured by ALICE in p-p collisions at 0.9, 2.36 and 7 TeV centre-of-mass energy 
have been presented. The measured values of the pseudorapidity density at all
energies are significantly larger than predicted by the current models,
except for PYTHIA tune ATLAS-CSC. All the event generators significantly 
underpredict the pseudorapidity density increase with the energy.
They also fail in reproducing the shape of the measured multiplicity distributions:
the disagreement is not concentrated in a single
region of the distribution and varies with the model.

\end{document}